\documentclass[preprints,review,accept,pdftex,oneauthor]{Definitions/mdpi} 
\usepackage{mathrsfs}
\usepackage{bm}
\firstpage{1} 
\pubvolume{1}
\issuenum{1}
\articlenumber{0}
\pubyear{2024}
\copyrightyear{2024}
\datereceived{ } 
\daterevised{ } 
\dateaccepted{ } 
\datepublished{ } 
\hreflink{https://doi.org/} 

\Title{Primordial black holes: formation, spin and type II}

\TitleCitation{Primordial black holes: formation, spin and type II}

\Author{Tomohiro Harada $^{1}$\orcidA{}}

\AuthorNames{Tomohiro Harada}

\AuthorCitation{Harada, T.}

\address{%
$^{1}$ \quad Department of Physics, Rikkyo University, Toshima, Tokyo 171-8501, Japan; harada@rikkyo.ac.jp}

\corres{Correspondence: harada@rikkyo.ac.jp}

\abstract{
Primordial black holes (PBHs) may have formed through the gravitational collapse of cosmological perturbations that were generated and stretched during the inflationary era, later entering the cosmological horizon during the decelerating phase, if their amplitudes were sufficiently large. In this review paper, we will briefly introduce the basic concept of PBHs and review the formation dynamics through this mechanism, the estimation of the initial spins of PBHs and the time evolution of type II fluctuations, with a focus on the radiation-dominated and (early) matter-dominated phases.
}


\begin{document}

%


\section{Introduction \label{sec:introduction}}

Black holes may have formed in the early Universe. This possibility was 
first considered by Zel'dovich and Novikov (1966)~\cite{Zeldovich:1967lct} 
and Hawking (1971)~\cite{10.1093/mnras/152.1.75}. 
These black holes have been termed primordial black holes (PBHs).
The observational relevance of PBHs to 
cosmology has been established by the subsequent 
work of Carr and Hawking (1974)~\cite{Carr:1974nx} and 
Carr (1975)~\cite{Carr:1975qj}.
See Refs.~\cite{Carr:2024nlv,Riotto:2024ayo} for the recent brief 
reviews of the history and the future of PBHs.

Although our Universe can be well approximated by a homogeneous and isotropic cosmological model -- the Friedmann-Lema\^{i}tre-Robertson-Walker (FLRW) solution -- on large scales, cosmological structures such as galaxy clusters, galaxies, planetary systems, and stars must still form. These structures arise from primordial fluctuations. Observations of anisotropies in the cosmic microwave background (CMB) radiation have revealed that the amplitude of these primordial fluctuations is generally very small. One of the most promising mechanisms to generate such fluctuations is inflation. In this scenario, quantum fluctuations provide the origin of the perturbations, strongly suggesting that the probability of generating large-amplitude perturbations would be small but finite. During inflation, these perturbations expand along with the inflationary expansion of the Universe. After inflation, when the Universe is reheated, sufficiently large-amplitude perturbations could eventually overcome cosmic expansion and collapse into black holes due to their self-gravity. Unlike the formation of typical astrophysical black holes, the formation of PBHs does not require stellar evolution.

The formation of PBHs through primordial fluctuations was initially proposed by Hawking (1971)~\cite{10.1093/mnras/152.1.75} and Carr and Hawking (1974)~\cite{Carr:1974nx} and further developed by Carr (1975)~\cite{Carr:1975qj}. This formation mechanism has since been confirmed through numerical simulations based on numerical relativity, pioneered by Nadezhin, Novikov and Polnarev (1978)~\cite{1978SvA....22..129N}. In this review paper, we will explore the recent developments in PBH formation studies and their implications for cosmology.

The motivation for PBH research is multi-faceted. Since PBHs can, in principle, be observed, they provide valuable information about the early Universe. In this sense, PBHs can be regarded as the fossils of the early Universe. Furthermore, they are also one of the most promising candidates for dark matter. Moreover, Hawking evaporation plays an important role in the study of PBHs. Hawking recognised PBHs as a unique laboratory for studying black hole evaporation when he discovered this phenomenon in 1974~\cite{Hawking:1974rv,Hawking:1975vcx}. Hawking evaporation is key to understanding quantum gravity, where the information loss problem has been extensively discussed. The details of this evaporation may depend on quantum gravity and high-energy physics. 
Recently, several gravitational wave observatories have become available as observational tools to study strongly gravitating but otherwise dark objects, following LIGO’s first direct detection of gravitational waves in 2015~\cite{LIGOScientific:2016aoc}. 
It has been realised that gravitational waves provide a unique means of probing 
PBHs~\cite{Sasaki:2018dmp}. 
This suggests that PBHs lie at the intersection of various growing fields of modern physics, such as cosmology, general relativity, gravitational waves, quantum gravity and high-energy physics.

As observational data accumulate, gravitational waves are becoming an increasingly important means of observing our Universe. More than 100 events have been observed by the LIGO-Virgo-KAGRA (LVK) collaboration, and many of these have been identified as binary black holes. Several groups proposed the possibility that these binary black holes may be of cosmological origin, as many of them are approximately 30 $M_{\odot}$, which is more massive than was typically expected based on the standard theory of stellar evolution before the direct observations~\cite{Sasaki:2016jop,Bird:2016dcv,Clesse:2016vqa}. As more observational data accumulate, information about not only the masses but also the spins of binary black holes is obtained for some events~\cite{LIGOScientific:2017bnn}, which may provide a new clue to the origin of these binary black holes. From the obtained mass function of binary black holes, a search for the population of PBHs was carried out~\cite{Franciolini:2021tla}. It has also been argued that the existence of subsolar candidates in LIGO data would be a smoking gun for PBHs because no astrophysical scenario has yet been established to form subsolar-mass black holes~\cite{Carr:2023tpt}. Most recently, the NANOGrav collaboration reported evidence for nanohertz gravitational waves~\cite{NANOGrav:2023gor}, which might be consistent with gravitational waves induced by scalar perturbations that could have produced the amount of PBHs responsible for the binary black holes observed by the LVK collaboration (e.g.~\cite{NANOGrav:2023hvm}).
The recent observations with the James Web Space Telescope have identified several very massive galaxies at high redshifts, which might indicate a tension with predictions in the standard $\Lambda$CDM model but can be explained if we include PBH clusters in the formation scenario~\cite{Liu:2022bvr,Hutsi:2022fzw,Yuan:2023bvh}.

I would also comment on the possibility of PBHs constituting all or a considerable fraction of dark matter.
We usually discuss this in terms of the fraction of PBHs to cold dark matter (CDM),
$f(M)=\Omega_{\rm PBH}(M)/\Omega_{\rm CDM}$, 
as a function of the mass of PBHs, $M$.
This fraction is observationally constrained very severely in some mass ranges but not in others. See Fig. 10 of Carr, Kohri, Sendouda and Yokoyama (2021)~\cite{Carr:2020gox} for an overview of the constraints.
Recent observational constraints indicate two intriguing windows for dark matter. 
One is $M \sim 10^{17}-10^{23}$ g, where $f(M)$ is virtually unconstrained. This implies that PBHs could account for all the 
CDM for this mass range. The other mass range is   
$M \sim 1-10^{3} M_{\odot}$, where $f(M)\lesssim 0.1$, which is of interest in the context of terrestrial gravitational wave observations. 
Although there is currently a mass window in which all the CDM might be explained by PBHs, a stricter constraint could be placed on this mass window in the near future. Even if this turns out to be the case, it does not imply that studies of PBHs are without value. In this context, I would like to quote a profound statement by Bernard Carr: 
``Indeed their study may place interesting constraints on the
physics relevant to these areas even if they never formed''~\cite{Carr:2003bj}.

The aim of this article is to briefly review theoretical studies of PBH formation from primordial fluctuations,
the recent estimation of the spins of PBHs and the 
formation of PBHs generated from the so-called type II configurations.
In Sec.~\ref{sec:basic_concept}, we present the basic concept of PBHs. We discuss their masses, the Hawking evaporation, the fraction of their contribution to the whole dark matter and the formation probability. In Sec.~\ref{sec:formation}, we discuss the formation of PBHs. We focus on the formation mechanism from fluctuations generated by inflation and present key ideas on this scenario. 
In Sec~\ref{sec:spin}, we discuss the spins of PBHs.
The spins of PBHs have often been regarded just as negligible until recently. 
We quantitatively discuss the spins of PBHs just formed not only 
in a radiation-dominated era
but also in an (early) matter-dominated era. 
In Sec.~\ref{sec:type_II}, we discuss the initial configurations featured with their spatial geometry with throat structure, which is called type II, and the formation of PBHs from them based on the result of the recently conducted numerical simulations. In Sec.~\ref{sec:conclusions}, we conclude the paper.

\section{Basic concept of primordial black holes \label{sec:basic_concept}}

\subsection{Mass}
The striking feature of PBHs is that they can have a large range of possible mass scales 
from $\sim 10^{-5}$ g to $\sim 10^{56}$ g depending 
on the formation scenarios at least in principle.
Although the mass of the PBH may depend on the scenario, 
we usually assume that it can be approximately given by the mass enclosed within 
the cosmological horizon or Hubble horizon at the formation time $t_{f}$ from the big bang as
\begin{eqnarray}
 M\simeq M_{H}(t_{f})\simeq \frac{c^{3}}{G}t_{f} \simeq
      1 M_{\odot}~\left(\frac{t_{f}}{10^{-5}~\mathrm{s}}\right), 
\label{eq:pbh_mass}
\end{eqnarray}
where $c$ and $G$ are the speed of light and the gravitational constant, respectively, 
for which the gravitational radius is given by 
\begin{equation}
 R_{g}\simeq 1~\mathrm{km}~\left(\frac{M}{M_{\odot}}\right). 
\end{equation}
So, we can say ``the smaller, the older''. Table~\ref{table:PBHmass}
shows the relation between the formation time and the initial PBH mass.
Nevertheless, we should keep a caveat in our mind that the mass of PBHs could be 
much smaller than the horizon mass in certain formation scenarios.

\begin{table}[htbp] 
\caption{The cosmological time from big bang and the initial mass of PBHs if they are formed then.
\label{table:PBHmass}}
\newcolumntype{C}{>{\centering\arraybackslash}X}
\begin{tabularx}{\textwidth}{LL}
\toprule
\textbf{Cosmological time}	& \textbf{Mass of PBHs} \\	
\midrule
$\sim 10^{-43}$ s [Planck time] & $\sim 10^{-5}$ g [Planck mass]\\
$\sim 10^{-23}$ s & $\sim 10^{15}$ g [Critical Mass] \\
$\sim 10^{-5}$ s [QCD crossover] & $\sim 10^{33}$ g [Solar mass] \\
$\sim 10^{12}$ s [Matter-radiation equality] & $\sim 10^{50}$ g \\
$\sim 10^{19}$ s [Present epoch] & $\sim 10^{56}$ g [Mass of the observable Universe] \\
\bottomrule
\end{tabularx}
\end{table}

The mass of PBHs may have changed in time after they formed.
As for the mass accretion, earlier works suggest that  
it does not so significantly affect the mass at least 
in radiation domination~\cite{Carr:1974nx}.
This was confirmed by numerical simulations (e.g.~\cite{Escriva:2019nsa}).
On the other hand, the Hawking evaporation is considered to decrease the mass of PBHs. 
The mass loss is considered to be virtually negligible if the mass of the PBH is much more massive than the critical mass $\sim 10^{15}$ g, which will be discussed below.

\subsection{Evaporation}
Based on quantum field theory in curved spacetimes, 
Hawking (1974)~\cite{Hawking:1974rv,Hawking:1975vcx} found  that
black holes emit black body radiation. The temperature $T_{H}$ of the black body,
which is called the Hawking temperature, is proportional to the 
surface gravity of the horizon and is given for the Schwarzschild black hole by
\begin{eqnarray}
T_{H}=\frac{\hbar c^{3}}{8\pi GMk}\simeq 100~\mathrm{MeV}~
 \left(\frac{M}{10^{15}\mathrm{g}}\right)^{-1},  
\end{eqnarray}
where $\hbar$ and $k$ are the reduced Planck constant and the Boltzmann constant, respectively.
This is called the Hawking evaporation. If we assume that the black hole loses its 
mass due to this radiation of quantum fields according to the Stefan-Boltzmann law, which is 
the so-called semi-classical approximation, we obtain 
\begin{equation}
\frac{dM}{dt}=
-\frac{g_{\rm eff}\hbar c^{4}}{15360\pi G^{2}M^{2}},
\end{equation}
where $g_{\rm eff}\sim 100$ is the effective degrees of freedom and the grey-body factor is 
neglected. This implies that the evaporation timescale $t_{\rm ev}$ is given by 
\begin{equation}
t_{\rm ev}\simeq \frac{G^{2}M^{3}}{g_{\rm eff}\hbar c^{4} }\simeq 10
~\mbox{Gyr}\left(\frac{M}{10^{15} \mbox{g}}\right)^{3}, 
\label{eq:evaporation_timescale} 
\end{equation}
after which the black hole of mass $M$ loses almost all of its mass. As the black hole decreases 
its mass, the temperate gets higher and higher and 
the evaporation timescale becomes shorter and shorter. Although the black hole 
decreases its mass very slowly in most of its life, it rapidly loses its remaining 
mass at its final moment, which is called a black hole explosion.
When it becomes as light as 
the Planck mass $m_{\rm Pl}\sim 10^{-5}$ g, the semi-classical approximation 
necessarily breaks down and the subsequent evolution of the black hole should greatly 
depend on quantum gravity.  

From Eq.~(\ref{eq:evaporation_timescale}), the critical mass is 
approximately given by $10^{15}$ g for which $t_{\rm ev}=t_{0}$ is 
satisfied, where $t_{0}$ is the age of the Universe. So, if $M\lesssim 10^{15}$ g, 
PBHs have dried up after the explosion until now, whether they leave Planck mass relics
or not. If $M\sim 10^{15}-10^{17}$ g, 
PBHs are currently emitting X rays and $\gamma$ rays
and can be observed through those emissions. If $M\gtrsim 10^{17}$ g,  
the evaporation is mostly negligible and 
the mass of the PBH remains almost constant until now. 
It should also be noted that very recently there has been a debate on 
a specific hypothetical scenario that assumes that the Hawking evaporation 
be strongly affected due to the 
back reaction effect even if the black hole is macroscopic~\cite{Dvali:2020wft}.

\subsection{Probability}
To discuss the formation, the fraction $\beta(M)$ of the Universe which goes 
into PBHs when the mass contained within the cosmological horizon is $M$, is often used.
This can also be regarded as the formation probability of PBHs.
If we consider PBHs formed in the radiation-dominated phase, PBHs act as nonrelativistic particles surrounded by relativistic particles. Therefore, 
the energy density of PBHs, $\rho_{\rm PBH}$, decays as $a^{-3}$, while 
the energy density of 
relativistic particles, 
$\rho_{\rm rad}$, 
decays as $a^{-4}$, where $a$ is the scale factor of the Universe. 
This implies $\rho_{\rm PBH}/\rho_{\rm rad}\propto a$, that is, 
PBHs are condensed during the radiation-dominated era in proportion to the scale factor.
Taking this concentration effect into account, 
$\beta(M)$ is related to the 
the current fraction of 
PBHs of mass $M$ to all the CDM
\begin{equation}
 f(M)=\left.\frac{\Omega_{\rm PBH}}{\Omega_{\rm CDM}}\right|_{t=t_{0}}
\end{equation}
through 
\begin{eqnarray}
 \beta(M)&\simeq & 2\times 10^{-18}\left(\frac{M}{10^{15}~\mathrm{g}}\right)^{1/2}f(M)
\label{eq:beta_f}
\end{eqnarray}
for $M\gtrsim 10^{15}$~\mbox{g}~\cite{Carr:1975qj,Carr:2020gox}. 
This formula can derived by the expansion law $a(t)\propto t^{1/2} $ for the flat 
Friedmann-Lema\^{i}tre-Robertson-Walker (FLRW) solution in radiation domination, the mass within the horizon at the matter-radiation equality $M_{\rm eq}\sim 10^{50}$ g and Eq.~(\ref{eq:pbh_mass}). This equation is important because it connects the theory and the observation of PBHs.
Thus, for example, only a tiny probability $\beta\simeq 2\times 10^{-17}$ is enough to explain all of the dark matter if $M=10^{17}$ g, while $\beta\simeq 2\times 10^{-8}$ is necessary for that if $M=30 M_{\odot}$.

For $M\gtrsim 10^{15}$ g, Eq.~(\ref{eq:beta_f}) combined with the observational constraint on $f(M)$ gives the observational constraint on $\beta(M)$. As for $M\lesssim 10^{15}$ g, PBHs have evaporated away until now. However, the evaporation of PBHs may spoil big bang nucleosynthesis for $M\sim 10^{10}-10^{13}$ g if $\beta(M)$ is too large. Such a consideration gives the observational constraint on $\beta(M)$ for $M\gtrsim 10^{-5}$ g, where the lower limit is usually assumed to be the Planck mass. See Fig. 18 
of Ref.~\cite{Carr:2020gox} for more details.

\section{Formation \label{sec:formation}}

\subsection{Overview}

The serious study of PBH formation dates back to Carr (1975)~\cite{Carr:1975qj}. One of its aims is to theoretically predict $\beta(M)$
and, therefore, $f(M)$ and other observables of PBHs for a given cosmological scenario. 
We can obtain the information of the early Universe from observational data on PBHs only through these studies.
Apart from such an observational motivation, through the study of formation we can understand physics in PBH formation and investigate new phenomena and/or 
new physics in highly nonlinear general relativistic 
dynamics and high-energy physics. 
There have been a lot of scenarios proposed for PBH formation.
One of the most standard ones is the direct collapse of a large amplitude of  
primordial perturbations generated 
by inflation~\cite{Carr:1993aq,Carr:1994ar}. 
See also Ref.~\cite{Sasaki:2018dmp} for a recent review on this scenario.
This can be considered as `inevitable'  
as the inflationary cosmology has been regarded as 
an essential part of standard cosmology. 
Among the other alternative scenarios, domain wall collapse, bubble nucleation, collapse of string networks and phase transitions 
have attracted more attention than others. 
Hereafter, we will focus on the formation scenario from 
fluctuations generated by inflation. 
Studies on this scenario have developed more than on other scenarios.
The key ideas to this scenario is the following: fluctuations generated by inflation, long-wavelength solutions, formation threshold, black hole 
critical behaviour, dependence on the equation of state (EOS) and statistics on the abundance estimate. We will briefly view these topics below.
We will hereafter use the units in which $c=G=1$ over this paper unless they are explicitly given.

\subsection{Fluctuations generated by inflation}

The most striking feature of inflation is that it can not only solve the flatness problem and the horizon problem but also 
provide the mechanism to generate fluctuations through quantum effects.
Those fluctuations seed structure formation of different scales in our Universe
and provide anisotropies in CMB currently observed with high accuracy. 

\begin{figure}[htbp]
 \begin{center}
\includegraphics[width=0.8\textwidth]{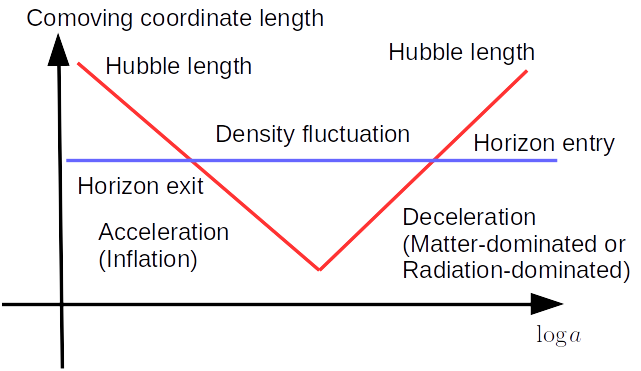}
\caption{Schematic figure for the evolution of the scales of fluctuations after generated by inflation.
The quantum fluctuations generated by inflation are stretched to the length scales much larger than the Hubble horizon scale $cH^{-1}$ and get classical. 
In the decelerated phase after inflation, 
the scale of the fluctuation expands slower than the Hubble horizon scale.
The time when the fluctuation scale gets as large as the Hubble scale is 
called the horizon entry of the perturbation. After the horizon entry, the perturbation 
can collapse to a black hole leading to PBH formation 
if its amplitude is sufficiently large.
\label{fg:horizon_entry}}
\end{center}
\end{figure}

The evolution of fluctuations in the inflationary cosmology is schematically illustrated 
in Fig.~\ref{fg:horizon_entry}. 
The quantum fluctuations generated by inflation are stretched to the length scales much larger than 
the Hubble horizon scale $cH^{-1}$ because the fluctuation scale, which is 
proportional to $a$,  
expands faster than 
the Hubble horizon scale, which is proportional to $t$, 
due to the accelerated expansion. 
This process is also considered to remove quantum coherence from the fluctuations. 
The generated classical fluctuations are further stretched away compared to the Hubble length as long as the inflationary phase continues. After the inflation ends, the expansion of the Universe begins to be decelerated.
There is a possibility that the Universe might experience the early matter-dominated phase due to the harmonic oscillation of an inflaton field. 
In any case, the Universe eventually gets dominated by the radiation field, tightly coupled relativistic particles in an almost complete thermal equilibrium state.  
This process is called reheating.
If we consider the scale of the fluctuation much larger than the Hubble horizon 
in the decelerated Universe, 
the scale of the fluctuation, 
which is proportional to $a$, expands slower than 
the Hubble horizon scale, which is proportional to $t$.
The time when the fluctuation scale gets as large as the Hubble scale is 
called the horizon entry of the perturbation.
Although the inflationary cosmology has been becoming an essential part of the standard cosmology,
there has been no standard inflation model until now. There are lots of inflation models, each of which gives the power spectrum $P_{\zeta}(k)$ and the other statistics of the 
curvature perturbations $\zeta$. 
See Ref.~\cite{Sasaki:2018dmp} for details of different inflation models 
in the context of PBH formation.

\subsection{Large-amplitude long-wavelength solutions}

As discussed above, fluctuations are generated and stretched to super-horizon scales by inflation. According to the standard cosmological scenario, the 
Universe must get radiation-dominated and turn to be decelerated 
after inflation. Then, the scale of the fluctuations gets as large as the horizon scale, 
or in other words, the fluctuations re-enter the horizon. Then, if the amplitude of 
perturbation is nonlinearly large, the fluctuation can collapse to a black hole in the 
radiation-dominated era. 
This scenario has been conventionally studied for many years. In some scenarios, 
there can be an early matter-dominated era before the standard late one. 
In this case, we also need to think the collapse of the fluctuation in the 
early matter-dominated era.

\begin{figure}[htbp]
\begin{center}
 \includegraphics[width=0.8\textwidth]{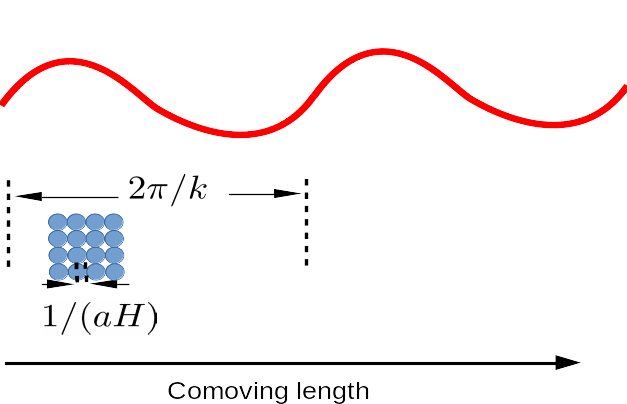}
\caption{Schematic figure of long-wavelength solutions. 
The long-wavelength solutions are obtained under the assumption that the comoving length scale $2\pi/k$ of the perturbation in consideration is much longer than the Hubble horizon length $1/(aH)$, which gives the size of Hubble patches denoted with small blue disks.
 \label{fg:long_wavelength}}
\end{center} 
\end{figure}

Based on the above scenarios, we discuss the fluctuations of super-horizon scales in the decelerated expansion after inflation.
This situation is schematically illustrated in Fig.~\ref{fg:long_wavelength}.
As we have already seen, 
in order to discuss PBH formation in radiation domination, 
it is necessary to deal with a nonlinearly large amplitude of perturbation of 
super-horizon scales. To obtain such perturbations, 
we solve the Einstein equation by a series expansion 
assuming that the spatial derivative is regarded as 
much smaller than the time derivative. This scheme is called 
gradient expansion and the solutions obtained by this scheme
are called long-wavelength solutions.

First, we make 
the standard $3+1$ decomposition of the 
spacetime. 
We foliate four-dimensional spacetimes with constant $t$
spacelike hypersurfaces $\Sigma_{t}$. Let $x^{i}$ and 
$\gamma_{ij}(t,x^{k})$ be spatial coordinates on $\Sigma_{t}$
and the metric tensor on $\Sigma_{t}$, respectively, where
$i,j,k,\cdots$ run over 1,2 and 3.
Then, the line element in four dimensions can be written by 
\begin{equation}
 ds^{2}=-\alpha^{2}dt^{2}+\gamma_{ij}(dx^{i}+\beta^{i}dt)
(dx^{j}+\beta^{j}dt), 
\end{equation}
where  $\alpha$ and $\beta^{i}$ are functions of $t$ and $x^{i}$ 
and called the lapse function and the shift vector, respectively.
The functions $\alpha$ and $\beta^{i}$ 
specify the time slicing, i.e., 
how the spacetime is foliated with constant $t$ spacelike hypersurfaces, 
and the coordinate threading, i.e., how the constant $x^{i}$
world lines cross the constant $t$ hypersurfaces, respectively.
We can choose the lapse and the shift according to the purposes
so that we can fix four degrees of the gauge freedom in the Einstein 
equation. Using this decomposition, we can reformulate the Einstein equation 
in the form of the second-order time development, i.e., the Cauchy problem.
See e.g. Refs.~\cite{Wald:1984rg,Poisson:2009pwt} for more details.

In the next step towards the long-wavelength solutions, 
we take the following procedures~\cite{1972PThPh..48.1503T,Shibata:1999zs, Lyth:2004gb, Polnarev:2006aa,Harada:2015yda}.
In our situation, it is more useful to further 
decompose $\gamma_{ij}$ in the form
\begin{equation}
 \gamma_{ij}=e^{2\zeta}a^{2}(t)\tilde{\gamma}_{ij},
\end{equation}
where $\tilde{\gamma}_{ij}$ and $\zeta$ are both  
the functions of $t$ and $x^{i}$ so that 
$\det (\tilde{\gamma}_{ij})=\det (\eta_{ij})$ should hold, 
where $\eta_{ij}$ is the time-independent metric of the flat 3-space,  
while $a(t)$ is the scale factor of the reference flat FLRW spacetime.
Thus, we have the line element in general spacetimes in four dimensions
in the following $3+1$ form:
\begin{equation}
 ds^{2}=-\alpha^{2}dt^{2}+e^{2\zeta}
a^{2}(t)\tilde{\gamma}_{ij}(dx^{i}+\beta^{i}dt)
(dx^{j}+\beta^{j}dt).
\end{equation}
We call this the cosmological $3+1$ conformal decomposition.

Then, we assume that the scale of the perturbation is much larger than 
the Hubble horizon scale, i.e., that 
\begin{equation}
 \epsilon:=\frac{k}{aH}
\end{equation} 
is much smaller than unity,where $k$ denotes the comoving wave number corresponding to the comoving scale of the perturbation.
Applying the gradient expansion in powers of $\epsilon$ for the Einstein equation, we can obtain growing-mode solutions with an assumption that the 
zeroth-order solutions in powers of $\epsilon$ takes the following form:
\begin{equation}
 ds^{2}=-dt^{2}+e^{2\zeta_{0}({\bm x})}
a^{2}(t)\eta_{ij}dx^{i}dx^{j},
\end{equation}
where $\zeta_{0}({\bm x})$ is
the zeroth-order part of $\zeta$, which is a function 
solely of ${\bm x}=(x^{1},x^{2},x^{3})$ and identified with primordial curvature perturbation.
Note that this assumption is compatible with the comoving slice, the constant-mean-curvature (CMC) slice and the uniform-density slice but not with the conformal Newtonian gauge condition, which is often used in cosmology~\footnote{As is well known, in the conformal Newtonian gauge, the density perturbation in the linear order approaches a time-independent function of ${\bm x}$ in the super-horizon limit, while it approaches zero in the same limit in the cosmological long-wavelength solutions 
formulated in Refs.~\cite{Shibata:1999zs, Lyth:2004gb, Polnarev:2006aa}. We do not exclude the existence of other formulations that are compatible with the conformal Newtonian gauge.}.
The higher-order terms of the solutions are obtained in terms of the function $\zeta_{0}({\bm x})$
if we impose the 
appropriate gauge conditions. In other words, $\zeta_{0}({\bm x})$, which is not suppressed due to the long-wavelength scheme, 
 generates the long-wavelength solutions.
For example, the density perturbation $\delta$ is $O(\epsilon^{2})$ and is given by  
\begin{equation}
 \delta_{\rm CMC}\simeq -\frac{4}{3a^{2}H^{2}}e^{-\frac{5\zeta_{0}}{2}}\Delta_{\rm flat}e^{\frac{\zeta_{0}}{2}} 
\end{equation}
in the CMC slice.

If we apply this formulation to the spherically symmetric spacetime, 
where the zeroth order metric can be written in the following form 
\begin{equation}
 ds^{2}=-dt^{2}+e^{2\zeta_{0}(r)}a^{2}(t)[dr^{2}+r^{2}(d\theta^{2}+\sin^{2}\theta d\phi^{2})],
\label{eq:0th-order_solution}
\end{equation}
in the conformally flat coordinates, 
we can show that this is equivalent to the asymptotically quasihomogeneous solutions developed in Ref.~\cite{Polnarev:2006aa} in the Misner-Sharp formulation, where 
the comoving slice and the comoving thread are adopted, and the explicit transformation between them is given in Ref.~\cite{Harada:2015yda}.

\subsection{Formation threshold in radiation domination}

If the amplitude of the perturbation, which is generated by inflation and enters the horizon in the radiation-dominated era, 
is sufficiently large, it will directly collapse to a black hole. Since the linear perturbation in the radiation-dominated era 
does not significantly grow, such a perturbation must be nonlinearly large. This prevents us from accessing the full dynamics of PBH formation with analytical methods. Only full numerical relativity simulations can accurately describe 
the general relativistic dynamics of PBH formation, 
which has been pioneered by Nadezhin, Novikov and 
Polnarev (1978)~\cite{1978SvA....22..129N}.
In fact, it has been established by several analytical and numerical works
such as Refs.~\cite{Carr:1975qj,1978SvA....22..129N}
that PBHs really form in this scenario.

This scenario naturally implies that there exists a threshold for PBH formation. 
Carr (1975) derived the threshold $\delta_{\rm th}\sim 1/3$ for radiation domination 
according to the Jeans scale argument
in terms of $\delta_{H}$, the density perturbation at the horizon entry of the perturbation. 
In numerical relativity, the threshold value is obtained as $\delta_{\rm th}\sim 0.45$ in terms of the density perturbation averaged in the comoving slice 
over $0<r<r_{0}$, where $r_{0}$ is the radius of the overdense region,
at the horizon entry of the overdense region, 
where $\delta_{H}$ is regarded as that in the nontrivial 
lowest order of the long-wavelength expansion scheme
\cite{Musco:2004ak,Musco:2008hv,Musco:2012au,Harada:2015yda}.
Harada, Yoo and Kohri (2013)~\cite{Harada:2013epa} 
refined Carr's argument from a general relativistic point of view 
and analytically 
derived the threshold $\delta_{\rm th}\simeq 0.41$ in the comoving slice.
Although the averaged density perturbation $\delta_{H}$ might look
straightforward to interpret,
it has difficulty in its interpretation. This is because if we calculate the 
averaged density perturbation at the horizon entry using the nontrivial lowest 
order of the solutions, it cannot be a real physical value, as the latter can 
only be obtained after the full numerical simulation and is not necessarily useful.

Shibata and Sasaki (1999)~\cite{Shibata:1999zs} 
defined a compaction function in the CMC slice. Although 
it was intended to equal to the ratio of the excess in the Misner-Sharp 
mass, or equivalently the Kodama mass, to the areal radius,
it is not equal to that but 
\begin{equation}
 {\cal C}_{\rm SS}\approx \frac{1}{2}\left[1-\left(1+r\zeta'\right)^{2}\right],
\end{equation}
where the weak equality denotes the equality in the long-wavelength limit and 
we have omitted the subscript $0$ in 
$\zeta$ according to the 
convention~\cite{Harada:2015yda,Harada:2023ffo}.
This has been shown to have a geometrical origin as a compactness function in the 
static spacetime obtained by removing the scale factor in the 
long-wavelength limit~\cite{Harada:2024trx}.
It is proportional to the ratio of the Misner-Sharp mass excess to 
the areal radius in the comoving slice as
\begin{equation}
 {\cal C}_{\rm SS}(r)\approx \frac{3}{4}C_{\rm com}(r),
\end{equation}
where 
\begin{equation}
 C_{\rm com}:=\frac{2\delta M_{\rm com}}{R}
\label{eq:C_com}
\end{equation}
with  
$\delta M_{\rm com}$ being the Misner-Sharp mass excess in the comoving slice. 
Note the factor of 2 on the right-hand side of Eq.~(\ref{eq:C_com}).
The threshold value is $\sim 0.4$ in terms of its maximum value of the Shibata-Sasaki 
compaction function ${\cal C}_{\rm SS}(r)$
in the long-wavelength limit~\cite{Shibata:1999zs,Musco:2012au,Harada:2015yda}.
This is more straightforward than the averaged density perturbation 
because for the compaction function description, we only have to take 
the long-wavelength limit of the solutions, where 
all the higher-order contributions naturally disappear. 
This is probably why people have favoured to use the compaction function.
On the other hand, this threshold can be transformed to 
$\sim 0.5$ in terms of the density perturbation 
in the comoving slice averaged over the ball $0<r<r_{m}$ at its horizon entry, 
where $r_{m}$ is the radius at which ${\cal C}_{\rm SS}$ takes the maximum~\cite{Musco:2018rwt}
again with a caveat that this value is only correct in the nontrivial lowest order of 
the long-wavelength expansion.
The averaging over $0<r<r_{m}$ is more consistent and applicable
than over the overdense ball $0<r<r_{0}$ as 
the definition of $r_{0}$ is problematic for density 
perturbations without an underdense region.

In the last decade, great effort has been paid to 
reveal the profile dependence of the threshold~\cite{Nakama:2013ica,Musco:2018rwt,Escriva:2019phb}. 
In particular, Escriv\`{a}, Sheth and Germani (2020)~\cite{Escriva:2019phb} found
a universal threshold
\begin{equation*}
 \bar{C}_{\rm com} \simeq \frac{2}{5}
\end{equation*}
in terms of $ \bar{C}_{\rm com}$, the spatial average of $C_{\rm com}$ over $0<r<r_{m}$, where $C_{\rm com}(r)$ takes a maximum at $r=r_{m}$. 
This holds within $2\%$ accuracy over different profiles they surveyed.
This gives a new threshold condition that uses the maximum $C_{\rm com}(r_{m})$ and 
its second-order derivative $C''_{\rm com}(r_{m})$.
See Ref.~\cite{Escriva:2019phb} for details.
More recently, Ianniccari, Iovino, Kehagias, Perrone and Riotto~\cite{Ianniccari:2024ltb} reported some numerical coincidence between this threshold and that for the existence 
of a circular photon orbit.  

There is a possibility that isocurvature perturbations may be produced depending on the cosmological scenario. Yoo, Harada, Hirano, Okawa and Sasaki (2021)~\cite{Yoo:2021fxs}, through numerical relativity simulations by introducing a massless scalar field perturbation into the radiation-dominated Universe, demonstrated that isocurvature perturbations can also produce PBHs in the radiation-dominated era. As isocurvature perturbations can play important roles in various intriguing inflationary scenarios, further studies are anticipated to achieve a comprehensive understanding of the formation threshold from isocurvature.

\subsection{Softer equation of state \label{subsec:PBH_soft}}

There remain a lot of theoretical possibilities for the thermal history of the Universe. 
In addition to the standard phases of radiation dominance and subsequent 
matter dominance, there may be early matter-dominated phase and/or 
phase transition or crossover phase in which the equation of state of the 
dominant component of the Universe is significantly softer than 
the radiation fluid.
Although PBH formation has been conventionally studied in the radiation-dominated phase, the scenarios in other phases with such significantly soft equations of state will also be important. This is because as we will see below, $\beta(M)$ strongly depends
on the threshold, which is smaller for the softer equation of state and, hence, 
PBH formation can be strongly enhanced so that PBHs formed in this phase
may dominate those formed in the radiation dominated era,
provided that the power spectrum density perturbation is nearly scale-invariant.

\begin{figure}[htbp]
\begin{center}
 \includegraphics[width=0.7\textwidth]{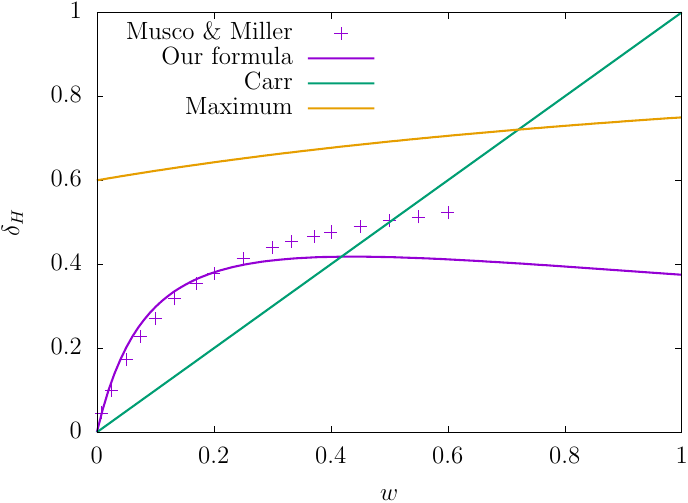}
\caption{EOS dependence of the PBH formation threshold.
The horizontal axis is the parameter $w$ for the EOS $p=w\rho$, while the vertical axis is the threshold value in terms of $\delta_{H}$, the density perturbation over the 
overdense region, $0<r<r_{0}$, at its horizon entry in the comoving slice. 
The crosses show the numerical result obtained in Ref.~\cite{Musco:2012au}. The purple curve is the plot of Eq.~(\ref{eq:HYK_formula}), which shows an agreement with the numerical result within $\sim 20\%$, while the green curve shows Carr's formula $\delta_{H}\simeq w$. See Ref.~\cite{Harada:2013epa} for more details.
\label{fg:delta_c_w_cut}} 
\end{center}
\end{figure}

This scenario was already suggested by Carr (1975)~\cite{Carr:1975qj}, where 
the threshold density perturbation was estimated to $\delta_{H}\sim w$ for 
the linear EOS $p=w\rho$ using the simple application of the Jeans criterion 
in Newtonian gravity. 
Harada, Yoo and Kohri (2013)~\cite{Harada:2013epa} 
refined it by comparing the free-fall time and the sound-crossing time 
in a simplified toy model of the fully general relativistic spacetime 
considering gauge difference to find the threshold value  
\begin{equation}
 \delta_{H}\simeq \frac{3(1+w)}{5+3w}\sin^{2}\left(\frac{\pi\sqrt{w}}{1+3w}\right)
\label{eq:HYK_formula}
\end{equation}
in the comoving slice.
This analytical expression showed a good agreement with the 
results of numerical relativity simulation~\cite{Musco:2012au,Escriva:2020tak}
for $0< w\le 1/3$ as seen in Fig.~\ref{fg:delta_c_w_cut}.
These studies showed that the threshold for the EOS $p=w\rho$ is an increasing function of $w$ for $0<w\le 1/3$ and approaches 0 as $w\to 0$.
For example, for the QCD crossover, for which $w$ drops to $\sim 0.23$ 
from $1/3$~\cite{Byrnes:2018clq}, this suggests enhanced generation of PBHs.
Although $w$ is time-dependent in this case, 
the threshold has also been calculated based on the Jeans 
criterion~\cite{Papanikolaou:2022cvo}.
Subsequently, full numerical relativity simulations have revealed 
that the PBH formation will be 
enhanced by a factor of the order of 1000~\cite{Escriva:2022bwe,Musco:2023dak}.
See Ref.~\cite{Jedamzik:2024wtq} for a recent review on this subject.

Musco and Papanikolaou (2022)~\cite{Musco:2021sva} studied 
the effect of anisotropic stress, 
which is the deviation from the perfect fluid description, 
on PBH formation in radiation domination and found that if this
is large enough, it could lead to a significant variation of 
the abundance of PBHs.

\subsection{Matter domination \label{subsec:PBH_MD}}

If the argument in Sec.~\ref{subsec:PBH_soft}
for the EOS $p=w\rho$ applied to $w\approx 0$ for the matter-dominated phase, 
one might consider that PBHs could be overproduced.
However, this cannot be correct 
because the realistic physical 
system is highly nonspherical and the deviation from spherical symmetry will grow during the collapse in matter domination. 
Not only after the standard matter-radiation equality time but also in a  
possible early matter-dominated phase,
which may naturally occur in the preheating process or in the strong phase transition 
in the Universe, 
PBH formation will be enhanced. This is one of the interesting epochs to study in the 
context of PBH formation. 
Since the condition for PBH formation is not solely determined by the pressure gradient force, we would need a totally different treatment for this phase 
from that for the radiation-dominated phase.  

\begin{figure}[htbp]
\begin{center}
 \includegraphics[width=0.95\textwidth]{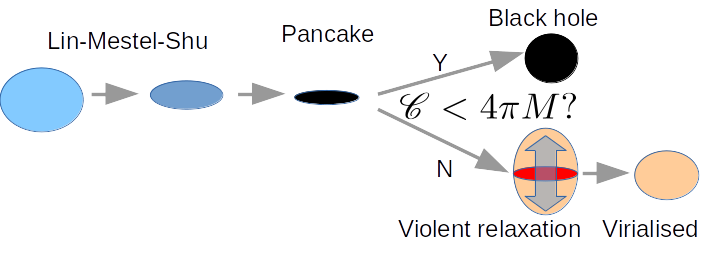}
\caption{Schematic illustration of the scenario on the anisotropic effect 
in the formation of PBHs in matter domination.
${\mathscr C}$ denotes the hoop
of the surface of the gravitating object of mass $M$.
The scenario is the following. The collapse of a dust ball is unstable against nonspherical perturbations towards the pancake. Whether or not a horizon forms around the pancake is subject to the hoop conjecture. If the condition ${\mathscr C}\lesssim 4\pi M$ is satisfied, a black hole forms. Otherwise the collapse leads to virialisation by acquiring velocity dispersion through violent relaxation.
See Ref.~\cite{Harada:2016mhb} for details. \label{fg:pancake_hoop}} 
\end{center}
\end{figure}

The PBH formation in matter domination has been pioneered 
by Khlopov and Polnarev (1980)~\cite{Khlopov:1980mg,1981SvA....25..406P,1982SvA....26..391P,1985SvPhU..28..213P}, 
where the effects of anisotropy and inhomogeneity are studied as 
obstruction of PBH formation process. 
Harada, Yoo, Kohri, Nakao and Jhingan (2016)~\cite{Harada:2016mhb} revisited the anisotropic effects
by combining the picture of pancake collapse of dark matter and the hoop conjecture by 
Thorne, which claims that black holes with horizons form when and
only when a mass $M$ gets compactified into a region whose
circumference in every direction is ${\mathscr C} \lesssim 4\pi M$
~\cite{Klauder:1972je,Misner:1973prb}.
This suppression due to the anisotropic effect is 
schematically shown in Fig.~\ref{fg:pancake_hoop}. 
They not only qualitatively reproduced 
the result of Ref.~\cite{Khlopov:1980mg,1982SvA....26..391P} 
but also updated the coefficient as
\begin{equation}
 \beta_{\rm aniso}(M)\simeq 0.05556 \sigma_{H}^{5}(M),
\end{equation}
where $\sigma_{H}(M)$ is the standard deviation of $\delta_{H}$ 
in the mass scale of $M$
and the Gaussian distribution for density perturbation is assumed.

Kopp, Hofmann and Weller (2010)~\cite{Kopp:2010sh} modelled spherical formation 
of PBHs in matter domination with the Lema\^{i}tre-Tolman-Bondi (LTB) solution 
and Harada and Jhingan (2015)~\cite{Harada:2015ewt} extended it to nonspherical formation with the Szekeres quasispherical solutions.
Kokubu, Kyutoku, Kohri and Harada (2018)~\cite{Kokubu:2018fxy} 
revisited the inhomogeneity effects by utilising the LTB solution and 
not only qualitatively 
reproduced the result of Ref.~\cite{Khlopov:1980mg,1982SvA....26..391P} 
but also updated the coefficient so that 
the additional suppression factor is given by 
\begin{equation}
 \beta_{\rm inhom}(M)\simeq 3.70\sigma_{H}^{3/2}(M)
\end{equation}
with the caveat that its physical effect largely depends on the 
assumption that black hole formation is prevented by the appearance of 
an extremely high-density region before the black hole horizon formation surrounding it.  

Harada, Kohri, Sasaki, Terada and Yoo (2022)~\cite{Harada:2022xjp}
showed that 
the effects of velocity dispersion that may have been 
generated in possible nonlinear growth of 
perturbation in the earlier phase can suppress PBH formation.
The effects of the angular momentum can play important roles
and suppress PBH formation for smaller $\sigma_{H}$~\cite{Harada:2017fjm}.
We will later discuss the angular momentum in the PBH formation 
in the matter-dominated era 
in the context of the initial spins of PBHs.

There is another intriguing approach to PBH formation during the early matter-dominated era. Such an era is thought to be naturally realised during the preheating phase following inflation, when an oscillating massive scalar field dominates the Universe. It is possible to gain further insights into PBH formation during this phase by directly studying the perturbations in the Universe dominated by a massive scalar field. This approach was explicitly proposed by Padilla, Hidalgo and Malik (2022)~\cite{Padilla:2021zgm,Padilla:2021uof,Hidalgo:2022yed}, and numerical relativity simulations have been conducted based on this method, involving a mixture of a massive scalar field and a massless scalar field~\cite{deJong:2021bbo,deJong:2023gsx,Aurrekoetxea:2023jwd}. This approach may involve highly fruitful physics, and further studies will be necessary for a comprehensive understanding of this scenario.

\subsection{Critical behaviour}

One of the great achievements of numerical relativity is the discovery of 
critical behaviour in gravitational collapse, which is also called black hole 
critical behaviour. This has been first discovered by 
Choptuik (1993)~\cite{Choptuik:1992jv}
in the spherically symmetric system of a massless scalar field 
and followed by Evans and Coleman (1994)~\cite{Evans:1994pj} 
in the spherical system of a radiation fluid. 
The phenomena were theoretically understood
in terms of the renormalisation group analysis 
by Koike, Hara and Adachi (1995)~\cite{Koike:1995jm}.
These early studies were all on asymptotically flat spacetimes
and whether they apply to PBH formation was not so trivial 
because of the different boundary conditions and the existence of the characteristic 
scale, the Hubble horizon length.
The critical behaviour in the PBH formation 
was discovered by Niemeyer and Jedamzik (1999)~\cite{Niemeyer:1999ak} 
but subsequently questioned~\cite{Hawke:2002rf}. 
Finally, Musco and Miller (2013)~\cite{Musco:2012au} beautifully 
confirmed the critical behaviour in PBH formation. 
The essence of the critical behaviour in the context of PBH formation is 
the following. Let us consider a one-parameter family of initial data 
of the Cauchy problem, 
for which we can choose the averaged density perturbation $\delta_{H}$
as the parameter of the family. The features of the critical behaviour 
do not depend on the choice of the one-parameter 
family of the initial data.
As we have already seen, there is a threshold value $\delta_{\rm th}$, 
beyond which a PBH forms. Then, the evolution of the initial data which has 
the critical value $\delta_{H}=\delta_{\rm th}$ approaches a 
particular member of continuously self-similar solutions, which is called a critical solution.
If $\delta_{H}$ is slightly above the critical value, 
we have PBH formation and 
the scaling law for the mass of the formed PBH, $M$, holds as follows:
\begin{equation}
 M\approx K M_{H}(\delta_{H}-\delta_{\rm th})^{\gamma},
\end{equation}
where $\gamma\simeq 0.36$ is called a critical exponent and 
$K$ is a nondimensional positive constant that depends on the shape of 
the initial perturbation, varying between 3 and 30 with a typical value $\sim 5$.
In fact, the critical behaviour, such as the critical solution and 
the critical exponent, 
does not depend on the choice of the one-parameter family of initial data, which 
is called universality. Note, however, that the critical solution and the critical 
exponent do depend on the matter field and the equation of state even 
if it is a perfect fluid~\cite{Maison:1995cc}.

In the context of PBH formation, if we assume that $\delta_{H}$ obeys
some reasonable statistical distribution, the critical behaviour implies that 
a tiny fraction of perturbations of mass scale $M_{H}$
can produce PBHs that are much smaller than $M_{H}$, while 
a large fraction still produce those of the order of $M_{H}$. This becomes important 
especially if we consider $\beta(M)$ for $M \sim 10^{16}-10^{17}$ g.
This is because the PBHs of the critical mass $\sim 10^{15}$ g
are severely constrained by observation through its X-ray or gamma-ray emission, 
while those of $\sim 10^{16}-10^{17}$ g are not. 
For example, the formation probability $\beta (M)$ 
with the horizon mass $M=10^{16}$ g is severely constrained 
by such a tiny fraction of PBHs of the critical mass $\sim 10^{15}$ g 
considerably smaller than $10^{16}$ g~\cite{Niemeyer:1997mt,Yokoyama:1998xd}. 

\subsection{Abundance estimation and statistics}
\label{subsec:abundance}

Even if we can identify the threshold condition and the classical dynamics of 
PBH formation and subsequent evolution, it is not sufficient to determine $\beta(M)$.
Clearly, we also need statistical properties of the perturbations. In inflationary cosmology,
the statistical properties of fluctuations generated by inflation,
such as the power spectrum $P_{\zeta}(k)$ and other 
statistical properties, can be predicted at least in principle 
if we fix the inflation model.

Carr (1975)~\cite{Carr:1975qj,Harada:2013epa} 
simply assumed that $\delta_{H}$ obeys a Gaussian distribution 
and obtained the following formula
\begin{equation}
\beta(M)\simeq 2\frac{1}{\sqrt{2\pi}\sigma_{H}(M)}\int^{\delta_{\rm max}}_{\delta_{\rm th}}
d\delta e^{-\frac{\delta^{2}}{2\sigma_{H}^{2}(M)}}
\simeq \sqrt{\frac{2}{\pi}}\frac{\sigma_{H}(M)}{\delta_{\rm th}}e^{-\frac{\delta_{\rm th}^{2}}{2\sigma_{H}^{2}(M)}}, 
\label{eq:Carr's_formula}
\end{equation}
where $\delta_{\rm max}=2/3$ and $\sigma_{H}(M)$ are 
the possible maximum value and the standard deviation of $\delta_{H}$ 
in the mass scale of $M$
and in the last approximation $\delta_{\rm th}\gg \sigma_{H}(M)$ is 
assumed~\footnote{The factor of two comes from the recommendation for structure formation by Press and Schechter (1975) ~\cite{Press:1973iz,Peebles:1994xt}. Note that whether or not this factor applies to PBH formation is highly nontrivial.}.
Since we need at least $\beta(M)\gtrsim 10^{-18}$ for PBHs to contribute to 
a considerable fraction of dark matter, we can conclude that only the 
Gaussian tail beyond $\sim 8\sigma_{H}$ is responsible for PBHs.
This implies that introducing 
small nonGaussianities can significantly enhance the formation of 
PBHs~\cite{Young:2013oia}. See Ref.~\cite{Pi:2024jwt} for a recent review on 
nonGaussianities in PBH formation and their link to induced gravitational waves.
If we consider PBHs formed in the radiation-dominated era
and assume $\delta_{\rm th}\sim 0.45$ or $\sim 0.5$, we need at least 
$\sigma_{H}(M)\gtrsim 0.05$ or $P_{\zeta}(k)\gtrsim 0.01$.
This is much larger than the observed value $P_{\zeta}(k)\simeq 10^{-10}$
for the CMB anisotropies, 
although it does not immediately exclude the considerable 
formation of PBHs because the scales of the PBHs and the CMB anisotropies 
are usually very different from each other. 

Although Carr's formula (\ref{eq:Carr's_formula}), which is also called the Press-Schechter approximation, 
is very useful, this is considered as a very rough approximation. One of the reasons is
that even if the curvature perturbation obeys a Gaussian distribution, the averaged 
density perturbation $\delta_{H}$ in the comoving slice cannot 
because it must be within a finite interval between the minimum $-1$ and 
the maximum $2/3$~\cite{Kopp:2010sh}. 
The other is that since PBHs will form only at very rare peaks of the perturbation,
peak theory should apply, which is known to give a physically reasonable prediction 
for galaxy formation~\cite{Bardeen:1985tr}.
The prediction of peak theory may be significantly different from that of Carr's formula
for the estimation of PBH abundance.
Currently, there are a few variations in the application of peak theory to the estimation of the PBH abundance,  which comes from theoretical ambiguity caused by incomplete understanding of nonlinear, nonspherical and multi-scale general relativistic dynamics~\cite{Yoo:2018kvb,Yoo:2020dkz,Germani:2019zez,Kitajima:2021fpq,Germani:2023ojx}.
See also Ref.~\cite{Young:2024jsu} for a recent review on this topic.

\section{Initial spins \label{sec:spin}}

If the black hole no hair conjecture holds in astrophysics, stationary black holes in vacuum should be 
well approximated by Kerr black holes which are characterised only by two parameters, the mass and the angular momentum. 
As we have already discussed, the mass of PBHs is approximately equal to that within the Hubble horizon at the time of formation. So, what determines the spin of PBHs?
If black holes of masses from several to several tens of solar masses are observed, 
it would be very difficult to distinguish between PBHs and astrophysical black holes.
Astrophysical black holes form in the final stage of evolution of massive stars. If we observe an isolated black hole, the only information of its own other than its mass is its spin. 
So, if the spins of PBHs are expected to be very different from those of astrophysical black holes, the observation of spin can be potentially decisive information to distinguish between the two populations. 
Here, we focus on the initial spins of PBHs.
However, it should be noted that 
the accretion and merger history after their formation could also greatly affect their spins depending on the scenarios~\cite{DeLuca:2020bjf} and, hence, the theoretically estimated values for the initial spins should be compared with observation with great care.

\subsection{Spins of primordial black holes formed in radiation domination \label{subsec:spin_RD}}

It was discussed that the formation process of PBHs in radiation domination is well approximated by spherically symmetric dynamics 
since Carr (1975)~\cite{Carr:1975qj}. 
Recent studies 
have revealed that this early argument is basically correct. 
Chiba and Yokoyama (2017)~\cite{Chiba:2017rvs} gave an upper bound $\sim 0.4$ on the root mean square of the nondimensional Kerr parameter $a_{*}=a/M$, where $a$ and $M$ are the Kerr parameter and the mass of the black hole. 
He and Suyama~(2019)~\cite{He:2019cdb} discussed the effect of 
angular momentum on the threshold.

Peak theory features more accurate quantitative analyses.
For PBHs formed in radiation domination, the threshold density perturbation $\delta_{\rm th}$ is of the order of the unity, which is considerably larger than its standard deviation. In other words, PBHs 
can form only at very rare peaks.  
Peak theory predicts that there is only very small deviation 
from spherical symmetry for such rare peaks.
This is very important to not only justify spherical symmetry assumption 
but also estimate the angular momentum of PBHs formed in radiation domination.
Based on the perturbative analysis based on peak theory,  
De Luca, Desjacques, Franciolini, Malhotra and Riotto (2019)~\cite{DeLuca:2019buf}
concluded that the root mean square of the nondimensional Kerr parameter $a_{*}$
is of the order of $10^{-2}$, while Harada, Yoo, Kohri, Koga and Monobe (2021)
estimated the root mean square of $a_{*}$ to be of the order of $10^{-3}$ and 
showed that a small fraction of PBHs of masses much smaller than the mass 
enclosed within the Hubble horizon, $M\ll M_{H}$, as a result of critical phenomena, can have much larger spins~\cite{Harada:2020pzb}.

\begin{figure}[H]
\begin{center}
\includegraphics[width=0.5\textwidth]{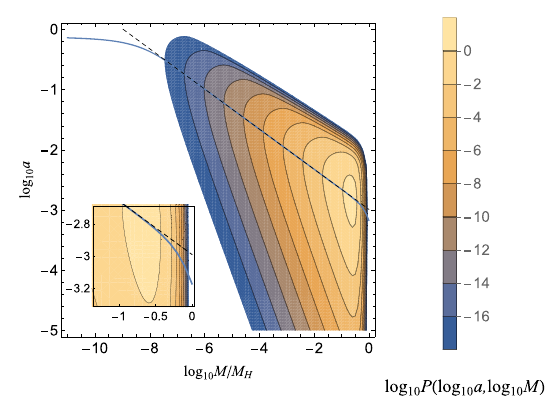}\\
\includegraphics[width=0.98\textwidth]{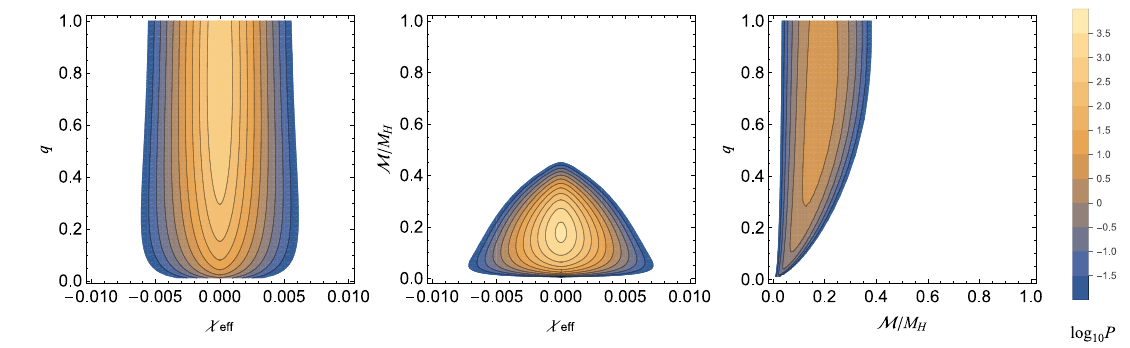}
\caption{
Contour maps of the distribution functions of $(M,a_{*})$ for isolated PBHs on the top panel, $M$ and $a_{*}$ 
are the mass and the nondimensional Kerr parameter, respectively, 
and 
$(\chi_{\rm eff},q)$, $(\chi_{\rm eff}, {\cal M})$ and $({\cal M},q)$ for binary PBHs from left to right on the bottom panels, 
where $\chi_{\rm eff}$, ${\cal M}$ and $q$ are
the effective spin parameter, the Chirp mass and the mass ratio of the binary PBHs, respectively. 
The definitions of $q$, ${\cal M}$ and $\chi_{\rm eff}$ are
$q:=M_{2}/M_{1}$, ${\cal M}:=(M_{1}M_{2})^{3/5}/(M_{1}+M_{2})^{1/5}$ 
and $\chi_{\rm eff}:=(a_{*1}\cos\theta_{1}+qa_{*2}\cos\theta_{2})/(1+q)$, respectively, 
where $M_{i}$, $a_{*i}$ and $\theta_{i}$ ($i=1,2$)
are individual masses, individual nondimensional Kerr parameters
and the angles of individual spins with respect to the orbital angular momentum,
respectively, with $M_{1}\ge M_{2}$ being assumed.
Taken from Ref.~\cite{Koga:2022bij}.
\label{fg:Koga_etal_2022_FIG2}}
\end{center} 
\end{figure}
Based on this analysis combined with peak theory, Koga, Harada, Tada, Yokoyama and Yoo (2022)~\cite{Koga:2022bij} calculated the distribution of the effective spin parameter $\chi_{\rm eff}$ of 
binary black holes, which 
is a well determined observable of the spins obtained 
from gravitational wave forms from inspiralling binaries.
Figure~\ref{fg:Koga_etal_2022_FIG2} 
shows the distribution functions of the parameters $a_{*}$ and $M$ of isolated PBHs
on the top panel 
and of the mass ratio $q:=M_{2}/M_{1}$, 
the Chirp mass ${\cal M}:=(M_{1}M_{2})^{3/5}/(M_{1}+M_{2})^{1/5}$ 
and $\chi_{\rm eff}:=(a_{*1}\cos\theta_{1}+qa_{*2}\cos\theta_{2})/(1+q)$ 
of binary PBHs on the bottom panels,
where $M_{i}$, $a_{*i}$ and $\theta_{i}$ ($i=1,2$)
are individual masses, individual nondimensional Kerr parameters
and the angles of individual spins with respect to the orbital angular momentum,
respectively, with $M_{1}\ge M_{2}$ being assumed.
Although there is a tendency for larger spins with  
smaller masses because of the critical phenomena, the probabilities for 
large values of the spin parameters are strongly suppressed. 
See Ref.~\cite{Koga:2022bij} for details.

In the above analyses, it was assumed that the power spectrum of curvature perturbation is nearly monochromatic. Recently, it has been revealed that if this assumption is relaxed, the root mean square of $a_{*}$ can be slightly larger for some set of broad power spectra but should still be bounded by the value of the order of $10^{-3}$~\cite{Banerjee:2024nkv}. 

\subsection{Spins of primordial black holes formed with a soft equation of state 
\label{subsec:spin_soft}}

It is interesting to ask how much the results in Sec.~\ref{subsec:spin_RD} depend 
on the properties of the matter fields in the cosmological phase when the PBHs formed. 
In particular, it was shown that the PBH production is significantly 
enhanced in the QCD crossover, where the effective value of $w$ drops 
from $1/3$ to $\sim 0.23$~\cite{Escriva:2022bwe,Musco:2023dak}. 
Saito, Harada, Koga and Yoo (2023)~\cite{Saito:2023fpt} showed that 
for a soft EOS parameterised by $p=w\rho$, the root mean square 
of $a_{*}$ is a decreasing function of $w$ and 
can be well fitted by the power law $\propto w^{-0.49}$. 
However, since the dependence is weak for $w\simeq 0.2-1/3$, 
the initial spins are only modestly enhanced 
for the QCD crossover such as to $\sim 0.003$ from $\sim 0.002$ for radiation. 
It also suggests that $a_{*}$ can be very large if $w\ll 1$, although 
the analysis in Ref.~\cite{Saito:2023fpt} is not well justified in the limit $w\to 0$,
where the treatment for matter domination should apply.

\subsection{Spins of primordial black holes formed in (early) matter domination
\label{subsec:spin_MD}}

\begin{figure}[htbp]
\begin{center}
{\includegraphics[width=0.5\textwidth]{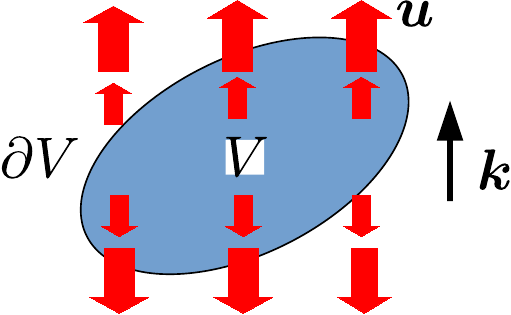}}
\caption{Schematic illustration of the first-order effect to generate angular momentum.
The blue region $V$ denotes the region that will collapse, which is approximated by an ellipsoid. The red arrows denote the velocity perturbation ${\bm u}$ with a wave number vector being parallel to the vertical direction. Since the ellipsoidal region $V$ is misaligned with the wave number ${\bm k}$ of the velocity perturbation, it carries 
nonvanishing angular momentum. The angular momentum can increase in time as the perturbation grows. 
This can also be understood through the effect of torque exerted on the boundary $\partial V$. 
See Ref.~\cite{Harada:2017fjm} for details. \label{fg:angular_momentum_MD}}
\end{center} 
\end{figure}

Since PBH formation will be enhanced in the (early) matter-dominated phase,
it is very important to predict the spins of them. As we have seen before, 
nonspherical effects may become important. In fact, 
Harada, Yoo, Kohri and Nakao (2017)~\cite{Harada:2017fjm}
investigated the effects of angular momentum for PBH formation in this phase.
Both the first-order and second-order effects can potentially play important roles. 
The first-order effect generates angular momentum through the nonsphericity of 
the region to collapse to a black hole, which is generally misaligned with a
mode wave number vector of the velocity perturbation.
This is schematically illustrated in Fig.~\ref{fg:angular_momentum_MD}, where 
the ellipsoidal region that will collapse is misaligned with the wave number
of the velocity perturbation denoted by the red arrows and, hence, it carries 
nonvanishing angular momentum.
The secondary effect also gives angular momentum through the 
coupling of two independent modes of linear perturbation, which is 
illustrated in Fig.~1 of Ref.~\cite{Harada:2017fjm}.

Although the dynamics of PBH formation in matter domination is expected to be very complicated, a perturbative calculation under certain working assumptions gives
\begin{equation}
 \langle a_{*}^{2} \rangle \sim  \sigma_{H}^{-1/2}, 
\end{equation}
where $a_{*}$ is the nondimensional spin parameter of the region to collapse and 
$\sigma_{H}$ is the standard deviation of $\delta_{H}$ at the the horizon entry.
Although the nondimensional numerical factor of the order of the unity 
on the right-hand side should be determined yet, 
this implies that most of PBHs have spins $a_{*}=O(1)$ 
if $\sigma_{H}\sim 0.1$. The angular momentum effects will strongly suppress 
PBH formation if $\sigma_{H}$ is even much smaller because of the Kerr bound $|a_{*}|\le 1$. These results have recently been updated based on peak theory~\cite{Saito:2024hlj}.

\subsection{Nonspherical simulation of PBH formation}

Although we have so far discussed the initial spins of PBHs, we have neglected 
nonspherical nonlinear general relativistic dynamics in the formation of black holes
for simplicity. It is clearly important to numerically simulate the nonspherical 
formation of PBHs and investigate how much 
the initially nonspherical initial data will affect the formation threshold 
and the initial spins of the produced PBHs. 
This is complementary to the perturbative analysis as
it can check the validity of assumptions made.

\begin{figure}[htbp]
\begin{center}
\begin{tabular}{cc}
\includegraphics[width=0.45\textwidth]{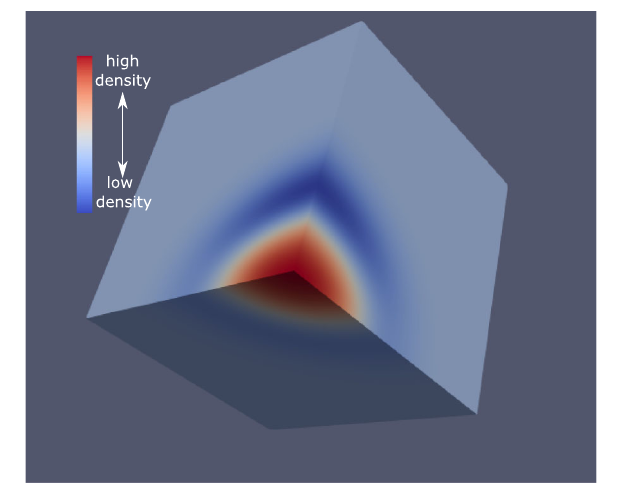}
&
\includegraphics[width=0.45\textwidth]{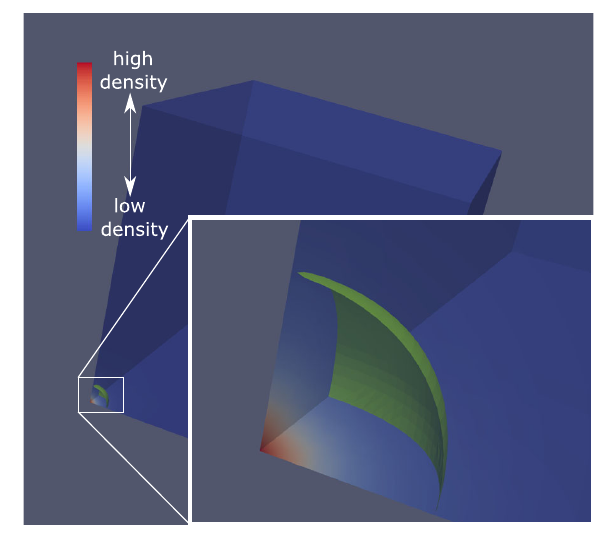}
\end{tabular}
\caption{The initial density perturbation on the left panel and the 
apparent horizon formation in 
the course of gravitational collapse on the right panel.
We can see that an apparent horizon forms in the central region, which is 
very small compared to the size of the whole computational domain.  
It implies that to resolve the apparent horizon for the near-critical collapse 
requires very high resolution at least near the centre.
Both taken from Ref.~\cite{Yoo:2020lmg} with permission.
\label{fg:YHO2020_FIG1_FIG3}}
\end{center} 
\end{figure}

Yoo, Harada and Okawa (2020)~\cite{Yoo:2020lmg} 
conducted 3D numerical simulation of nonspherical PBH formation 
in radiation domination based on numerical relativity for the first time. 
They prepared the long-wavelength solutions as 
initial data, for which initial nonsphericity is expected to be typically very small 
according to peak theory. However, to make the numerical results clearer, they 
put $\sim 10 \%$ nonsphericity in the initial data, which is much larger than 
the values expected from peak theory for PBH formation. 
They found that even such large 
nonsphericity changed the threshold of PBH formation only by $\sim 1 \%$.
See Fig.~\ref{fg:YHO2020_FIG1_FIG3} which is taken from Ref.~\cite{Yoo:2020lmg}
for the initial density perturbation on the left panel and the 
apparent horizon formation in 
the course of gravitational collapse on the right panel.   
We can see that in the simulation with the near-threshold value 
a very small apparent horizon forms at the very central region. 
This implies that we need very high resolution near the centre. 
This problem was attacked with a rescaled radial coordinate~\cite{Yoo:2020lmg}.  

Yoo (2024)~\cite{Yoo:2024lhp} conducted
further numerical relativity simulations 
for the EOS $p=w\rho$ with $w=0.2$ and $1/3$ with 
much higher resolution and accuracy.
For this purpose, not only the rescaled radial coordinate but also 
the multi-level mesh refinement scheme were adopted.
In such simulations, the estimation of the spin is not straightforward.
In this work, it was implemented as follows. 
For the Kerr black hole, we can write the spin parameter $a_{*}$ using the 
event horizon configuration through 
the following relation 
\begin{equation}
 a_{*}=\frac{\sqrt{4\pi A(d^{2}-\pi A)}}{d^{2}},
\label{eq:Kerr_spin_relation}
\end{equation}
where $d$ and $A$ are the equatorial circumference and 
the area, respectively.
Yoo (2024)~\cite{Yoo:2024lhp} 
estimated the spin of the PBH as $|a_{*}|\ll 0.1$ 
assuming Eq.~(\ref{eq:Kerr_spin_relation}) for the 
numerically found apparent horizon. 
This is consistent with the perturbative estimates discussed 
in Secs.~\ref{subsec:spin_RD} and \ref{subsec:spin_soft}.

As far as I am aware, no 3D numerical relativity simulations for nonspherical PBH formation in the (early) matter-dominated phase have been carried out using a fluid or dust description of the matter fields. In the scalar field approach to the early matter-dominated phase, De Jong, Aurrekoetxea, Lim and Fran\c{c}a (2023)~\cite{deJong:2023gsx} studied nonspherical PBH formation through 3D numerical relativity simulations and reported that the spin in the final stage of formation would be negligible. Although this result is highly suggestive, further systematic studies are anticipated in the near future.

\section{Type II perturbation and type B PBH \label{sec:type_II}}

\subsection{Introduction to type II perturbation}

In the FLRW solution, the geometry of the constant cosmological time $t$ spacelike hypersurface is 
given by a constant curvature space with a scale factor $a(t)$.
The metric can be written in the familiar form:
\begin{equation}
 ds^{2}=-dt^{2}+a^{2}(t)\left[\frac{1}{1-K\tilde{r}^2}d\tilde{r}^{2}+
\tilde{r}^{2}(d\theta^{2}+\sin^{2}\theta d\phi^{2})\right].
\label{eq:FLRW_metric}
\end{equation}
The spatial curvature $K$ can be zero, positive or negative. For $K\ne 0$, the geometry has a finite
curvature radius $1/\sqrt{|K|}$ up to the scale factor.
The positive curvature solution generally has particular time evolution in $a(t)$, 
which begins to expand from big bang, reaches
maximum expansion and collapses to big crunch, while for the zero and
negative cases it just continues expanding from big bang to infinity.
As is well known, for the positive-curvature FLRW spacetime, there is coordinate singularity
at $\tilde{r}=1/\sqrt{K}$ in the line element (\ref{eq:FLRW_metric}). We can go beyond this singularity
by introducing the radial coordinate $\chi$ in place of $\tilde{r}$ so that the line element can be written in the form:
\begin{equation}
 ds^{2}=-dt^{2}+a^{2}(t)[d\chi^{2}+\sin^{2}\chi (d\theta^{2}+\sin^{2}\theta d\phi^{2})],
\end{equation}
where $K$ is normalised to $1$ using the freedom in rescaling $a(t)$.
We can see that the coordinate singularity which was at $r=1$ in the $(t,r)$ coordinates
is now resolved at $\chi=\pi/2$ in the $(t,\chi)$ coordinates. 
Roughly speaking, 
a type II perturbation is the perturbation that covers the surface $\chi=\pi/2$, 
while a type I does not.

In the perturbation theory in the flat FLRW solution, there appear growing and 
decaying modes. We can usually neglect decaying modes in the early 
Universe if we are interested in subsequent structure formation including 
PBH formation. Then, we can expect that 
the regions with positive and negative density perturbations 
can be described by locally positive and negative 
curvature FLRW solutions, respectively. 
In fact, this is really true in the long-wavelength solutions because
Eq.~(\ref{eq:0th-order_solution}), which is the metric in the long-wavelength 
limit, can be rewritten in the following form~\cite{Polnarev:2006aa,Harada:2015yda}:
\begin{equation}
 ds^{2}=-dt^{2}+a^{2}(t)\left[\frac{1}{1-K(\tilde{r})\tilde{r}^2}d\tilde{r}^{2}+
\tilde{r}^{2}(d\theta^{2}+\sin^{2}\theta d\phi^{2})\right]
\end{equation}
in a new radial coordinate $\tilde{r}$.
Since type II perturbations need larger amplitudes of curvature perturbation, 
they are much rarer than type I in a standard probability distribution 
function of the curvature perturbation, whereas 
type II can be dominant in a particular inflationary 
scenario with large nonGaussianties~\cite{Escriva:2023uko}.

\subsection{Positive curvature region, type II perturbation and separate universe condition}

\begin{figure}[htbp]
 \begin{center}
 \includegraphics[width=0.5\textwidth]{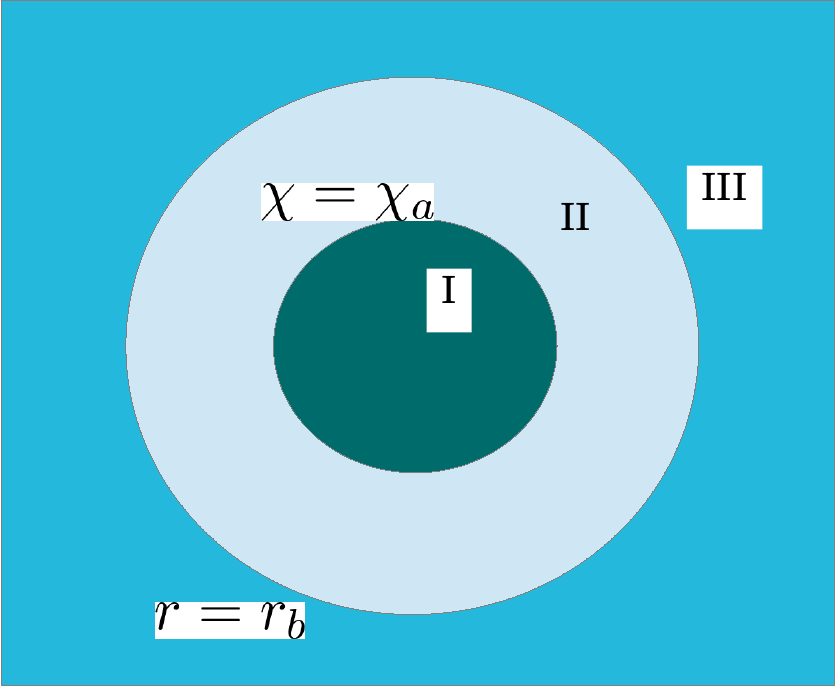}
\end{center}
\caption{The 3-zone model of the positive density perturbation. Regions I and III are described 
by the closed and flat FLRW solutions, while region II is an underdense matching layer. 
The spheres of $\chi=\chi_{a}$ and $r=r_{b}$ give 
the outer edge of region I and the inner edge of region III,
respectively. See Refs.~\cite{Kopp:2010sh,Harada:2013epa} for more details. \label{fg:3-zone_model}}
\end{figure}

As we have discussed, the region with a positive density perturbation can locally be 
well described by the positive-curvature FLRW solution at least in the long-wavelength limit.
For understanding the geometry and the dynamics of such a system, we introduce 
a toy model, which consists of 
a positive-curvature FLRW region surrounded by a flat FLRW 
region~\cite{Kopp:2010sh,Harada:2013epa}. We call this model the 3-zone model. 
Figure~\ref{fg:3-zone_model} gives its schematic figure. The line elements in regions I and 
III are written in the following forms, respectively, 
\begin{equation}
 ds^{2}=-dt^{2}+a^{2}_{\rm I}(t)\left[d\chi^{2}+\sin^{2}\chi (d\theta^{2}+\sin^{2}\theta d\phi^{2})\right]
\end{equation}
for $0<\chi<\chi_{a}$ and 
\begin{equation}
 ds^{2}=-dt^{2}+a^{2}_{\rm III}(t)\left[dr^{2}+r^{2}(d\theta^{2}+\sin^{2}\theta d\phi^{2})\right]
\end{equation}
for $r_{b}<r$, where $a_{\rm I}(t)$ and $a_{\rm III}(t)$ are the scale factors of the positive-curvature and flat FLRW solutions, 
while that in region II cannot be written in a 
simple form. Although this toy model is not only unrealistic but also 
difficult to justify as the whole evolution of the spacetime
unless we make strong assumptions on region II, it is still valid as the long-wavelength solution
and useful to understand the concept of the nonlinear perturbation. 
In this model, we have two independent physical length scales of region I, the curvature radius, which 
is normalised to $1$ in the above metric, and the size of the region, which is given by 
$\chi_{a}$. As $\chi_{a}$ is increased from $0$ to $\pi/2$, 
the comoving areal radius of region I, which is given by $\sin\chi_{a}$, monotonically 
increases from $0$ to $1$. However, as $\chi_{a}$ is further increased from $\pi/2$ to $\pi$,
the comoving areal radius turns to decrease from $1$ to $0$.  
This consideration implies the separate universe configuration for 
which $\chi_{a}=\pi$, which is notified in Ref.~\cite{Carr:1974nx}. 
Kopp, Hofmann and Weller (2010)~\cite{Kopp:2010sh} 
classified the configurations with $0<\chi_{a}<\pi/2$ and with $\pi/2<\chi_{a}<\pi$ 
into types I and II, respectively, while the marginal case is given by a 3-hemisphere $\chi_{a}=\pi/2$, where we neglect the contribution of region II.
 
\begin{figure}[htbp]
 \begin{center}
 \includegraphics[width=0.95\textwidth]{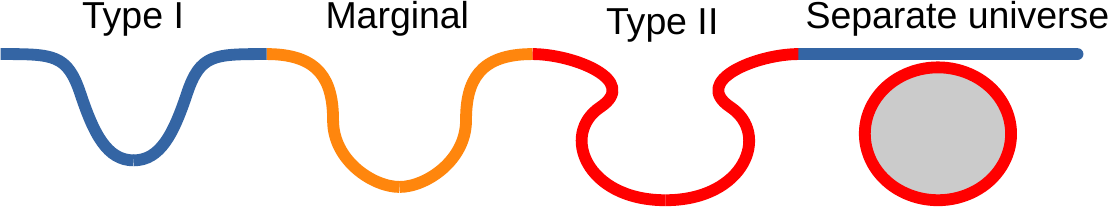}
\end{center}
\caption{Classification of spatial configurations of overdense perturbations. 
The type II configuration is with the throat structure in the spatial geometry, while the type I is not.
Note that the sequence does not correspond to the time evolution but to 
different initial curvature perturbations.
 \label{fg:separate_universes}}
\end{figure}

Apart from this toy model,
it is still true that
there should be two independent physical length scales to characterise the spatial geometry, the curvature radius and the size of the curved region
as schematically illustrated in 
Fig.~\ref{fg:separate_universes}.
It should be noted that this figure does not show the sequence of time evolution
but each configuration gives a set of initial data for each time development of 
perturbation.
It is physically interesting to think a nearly separate universe configuration, while 
a totally separate universe has nothing to do with our observable Universe.

\subsection{Type II perturbation and its time development} 

Here we review the sets of initial data in terms of types I and II based on 
the recent work by Uehara, Escriv\`{a}, Harada, Saito and Yoo 
(2024)~\cite{Uehara:2024yyp} 
for radiation domination. 
Here, the system is assumed to be spherically symmetric.
They first constructed the long-wavelength solutions with choosing a function $\zeta(r)$.
One of the functional forms they tried for $\zeta(r)$ is given by
\begin{equation}
 \zeta(r)=\mu e^{-(1/2)k^{2}r^{2}}W(r),
\label{eq:zeta_Gaussian}  
\end{equation}
where the constant $\mu$ gives the amplitude of perturbation and 
$W(r)$ is an appropriate window function chosen so that the Gaussian tail can be
eliminated and $\zeta$ can be smoothly matched to $0$ at the outer boundary of the numerical domain.
The areal radius $R$ and the Shibata-Sasaki compaction function ${\cal C}_{\rm SS}$
are plotted as functions of $r$ in Figs. 1(b) and 2(a) of~\cite{Uehara:2024yyp}, respectively, where we can see that for $\mu\gtrsim 1.4$, 
$R$ is no longer a monotonic function of the radial coordinate $r$, 
which implies that there is a throat and that ${\cal C}_{\rm SS}(r)$ has two 
peaks with the value of $1/2$ and a minimum in between.
See Ref.~\cite{Harada:2024trx,Uehara:2024yyp} for the account for 
this peculiar behaviour of the compaction function.
These features are essentially the same in matter domination 
as discussed in Ref.~\cite{Kopp:2010sh} using the LTB solution.

Then, the time development of those initial data was constructed
with a standard numerical relativity scheme
based on the Baumgarte-Shapiro-Shibata-Nakamura formalism but adjusted 
for the spherical formation of PBHs.
Figure 3 of \cite{Uehara:2024yyp} summarises 
the evolution of the spacetime for the long-wavelength solutions generated by 
the curvature perturbation given by Eq.~(\ref{eq:zeta_Gaussian}).
For $\mu=0.5$, the amplitude of perturbation is so small that it cannot collapse 
but disperses away. 
For $\mu=1.2$ and $1.8$, the the amplitude of perturbation is so large that it 
can collapse to a black hole.  
This suggests that the critical value of $\mu$ for the black hole formation
is between $0.5$ and $1.2$.

\subsection{Type B horizon structure}

In the asymptotically flat spacetimes, black holes are conventionally defined by event horizons.
However, in cosmological setting, event horizons are not necessarily useful
because the definition of an event horizon is teleological in the sense that it is only based on 
the infinite future and because the structure of infinities in cosmological spacetimes 
can be very different from asymptotically flat ones.
Furthermore, in the expanding Universe, there may be another type of horizons, 
i.e., a cosmological horizon. As we will see
below, we encounter the interplay of these two types of horizons for some class of 
PBH formation from the type II configuration. This implies that we need to 
distinguish between these two types of horizons. 
For these reasons, we adopt the notion of trapping horizons, a quasilocal
formulation of horizons~\cite{Hayward:1993wb,Hayward:1994bu}.  
This is akin to the notion of apparent horizons in the present setting.
Here, we define horizons using the pair of null expansions $\theta_{+}$ and $\theta_{-}$ 
along outgoing and ingoing radial null coordinates $\xi_{+}$ and $\xi_{-}$, respectively. 
Black hole horizons can be defined
with one vanishing and one negative null expansion 
so that any radial light ray there cannot expand its cross section. 
Similarly, cosmological horizons can be defined with one vanishing and 
one positive null expansion so that any radial light ray there cannot 
contract its cross section.
The black hole horizons and cosmological horizons are called 
future trapping horizons and past trapping horizons, respectively. 
In principle, we may also have a horizon with both null expansions vanishing, 
which is called a bifurcating trapping horizon~\cite{Maeda:2009tk}.
In Appendix~\ref{sec:trapping_horizon}, 
we will introduce trapping horizons in a more mathematical manner,
whereas the intuitive understanding discussed above is still useful.

\begin{figure}[htbp]
\begin{center}
\includegraphics[width=0.5\textwidth]{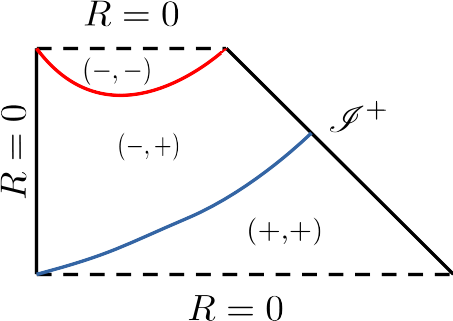}
\vspace{0.5cm}\\
\includegraphics[width=0.55\textwidth]{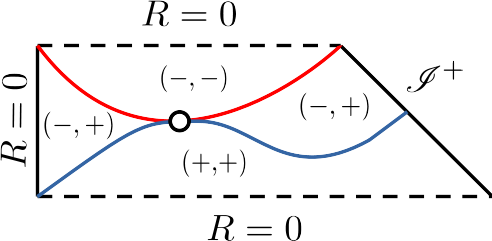}
\vspace{0.5cm}\\
\includegraphics[width=0.6\textwidth]{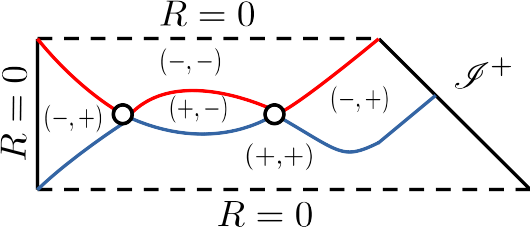}
\vspace{0.5cm}\\
\caption{Structures of trapping horizons and trapped regions 
inferred by numerical simulations. The structures indicated on the top, middle 
and bottom panels are called type A, marginal and type B, respectively. 
The signs of the radial null expansions $({\rm sign}(\theta_{-}),{\rm sign}(\theta_{+}))$ are shown for each region. The regions are divided by trapping horizons, where $\theta_{+}\theta_{-}=0$. The future and past trapping horizons are denoted by the red and blue curves, respectively. The bifurcating trapping horizons are denoted by open circles.
The black dashed lines denote spacetime singularities, while the black solid lines denote 
the regular centres and the null infinities. 
This figure is comparable to Fig. 7 of Ref.~\cite{Uehara:2024yyp}.
See text for more details.
\label{fg:horizon_structure}}
\end{center} 
\end{figure}

With the above terminology, we can now describe the structure of trapping horizons.
We assign $\xi_{+}$ and $\xi_{-}$ as the null coordinates being 
the standard ones in the far region which is asymptotic to 
the flat FLRW solution, which we assume to exist. 
There is a big difference in the structure of trapping horizons and trapped regions 
between $\mu=1.2$ and $\mu=1.8$.
We schematically plot the horizon structures inferred by numerical simulations in 
Fig.~\ref{fg:horizon_structure}.
For $\mu=1.2$, as shown on the top panel, 
there are a past trapped region and a future-trapped region that are disconnected from each other with the associated past and future trapping horizons being 
separate. We call this horizon structure type A.
For $\mu=1.8$, as shown on the bottom panel, 
the past and future trapped regions
have contacts at two points. The two points of intersection, 
which are in fact two 2-spheres,
correspond to bifurcating trapping horizons.
There appears an untrapped region enclosed by 
the future and past trapping horizons and 
the two bifurcating trapping horizons,
where $\theta_{+}<0$ and $\theta_{-}>0$. 
We call this horizon structure type B.

This numerical result for the perturbation
indicates that there appears the peculiar structure of trapping horizons and 
trapped regions, if $\mu$ is greater than some critical value. 
This was also true if we took another choice of $\zeta(r)$, so the above
features are at least general to some extent. 
So, we can regard this structure
as common for a sufficiently large amplitude of curvature perturbation of type II. 
We infer the existence of the marginal horizon structure 
between types A and B, which is shown on the middle panel of Fig.~\ref{fg:horizon_structure}. We can say that a type II perturbation does not always result in a type B
structure for radiation. 
This is due to the effect of pressure because it is known that for the dust case
a type II perturbation necessarily entails the horizon structure of type B. 
So, we can conclude that 
there are at least PBHs with structure of types I-A, II-A and II-B
in radiation domination.

It should be noted that the type B horizon structure has not been known in the gravitational collapse in asymptotically flat spacetimes and therefore very unique to PBH formation as long as the author is aware. For this reason, the result of the type II-B PBH formation is very intriguing in the context of the variety of black hole formation in general relativity. On the other hand, 
their relevance to observational cosmology is yet unclear.
 
\section{Conclusions \label{sec:conclusions}}

In the recent development of research on this subject, it has been revealed that PBHs lie at the intersection of various developing branches of modern physics. In this article, we review their basic concept, formation, spins and link to the type II configuration.

In Sec.~\ref{sec:basic_concept}, we present the basic concept of PBHs. The mass of PBHs is usually considered to be the mass scale within the cosmological horizon at the time of their formation. Mass accretion may significantly increase the PBH mass depending on the evolution scenario, whereas it has been shown to be negligible during the evolution in radiation domination. The mass of PBHs can be significantly reduced by Hawking evaporation, and they would have evaporated completely by now if they were lighter than the critical mass of approximately $10^{15}$ g, although the details of the evaporation process are still somewhat under debate. We can obtain constraints on the fraction $f(M)$ of PBHs of mass $M$ in relation to all dark matter through different observations. The fraction $f(M)$ can be transformed into the formation probability $\beta(M)$ depending on the cosmological evolutionary scenario. The standard cosmic history implies that a very small value of $\beta(M)$, as small as $\sim 10^{-17}$ for $M\sim 10^{17}$ g, can yield $f(M)=O(1)$, i.e., can explain all dark matter, because the PBHs' contribution to the energy of the Universe increases in proportion to the scale factor during the radiation-dominated phase.

In Sec.~\ref{sec:formation}, we discuss PBH formation. A detailed and precise understanding of PBH formation physics has become increasingly important. The basic question in this study is how to predict $\beta(M)$ and other observationally significant quantities from a given cosmological scenario. Focusing on PBH formation from fluctuations generated by inflation, the key terms are inflation models, long-wavelength solutions, thresholds, softer EOS, matter domination, critical behaviours and statistics.

In Sec~\ref{sec:spin}, we discuss the initial spins of PBHs. PBHs formed during radiation domination are unlikely to have large spins. Perturbative studies show that the nondimensional Kerr parameter of these PBHs is typically of the order of $10^{-3}$. In contrast, PBHs formed during matter domination can acquire large spins, at least initially. The effect of mass accretion after formation on the nondimensional Kerr parameter needs to be studied carefully. It is evident that numerical simulations based on numerical relativity should shed light on this problem.

In Sec.~\ref{sec:type_II}, we review the recently conducted numerical relativity simulations of the time evolution of type II perturbations during radiation domination. This is particularly relevant to a very rare perturbation peak or to a specific inflationary scenario. The resulting structures of trapping horizons can be classified into two types: one is standard for PBHs, and the other is very unique, featured with the crossing of trapping horizons as bifurcating trapping horizons. We refer to these as types A and B, respectively, while we can also discuss the marginal structure. The numerical simulations suggest that the evolution of type II perturbations can be classified into type A and type B based on their horizon structure. Thus, we may call them types II-A and II-B.

Finally, I must acknowledge that there are many interesting issues concerning the formation of PBHs that cannot even be mentioned in this article. The study of PBH formation not only requires a deep understanding of physical phenomena within known standard physics but also offers the opportunity to explore unknown new physics through PBHs. Both of these aspects will play important roles in the future of PBH formation studies.

\acknowledgments{
The main part of this article is based on exciting collaborations with 
Bernard Carr, 
Albert Escriv\`{a}, 
Jaume Garriga, 
Shin-Ichi Hirano,
Hayami Iizuka,
Sanjay Jhingan, 
Yasutaka Koga, 
Kazunori Kohri, 
Takafumi Kokubu, 
Koutaro Kyutoku,
Hideki Maeda, 
Takeru Monobe, 
Tomohiro Nakama, 
Ken-Ichi Nakao, 
Hirotada Okawa, 
A. G. Polnarev, 
Daiki Saito, 
Misao Sasaki, 
Yuichiro Tada, 
Takahiro Terada, 
Koichiro Uehara, 
Jun'ichi Yokoyama, 
Shuichiro Yokoyama 
and Chul-Moon Yoo.
The author thanks Cristiano Germani, Mohammad Ali Gorji 
and the other participants of the conference ``Barcelona Black Holes (BBH) I: Primordial Black Holes'', where this work was first presented, for helpful discussions and comments.
The author is also very grateful to Ilia Musco and 
Theodoros Papanikolaou for very helpful comments.
The author is grateful to 
CENTRA, Departamento de F\'{í}sica, Instituto Superior T\'{e}cnico -- IST at 
Universidade de Lisboa, and 
Niels Bohr International Academy at Niels Bohr Institute
for their hospitality during the writing of this manuscript.
}

\funding{This research was funded by JSPS KAKENHI Grant Numbers JP20H05853 and JP24K07027.}

\dataavailability{No data are associated with the manuscript.}

\conflictsofinterest{
The authors declare no conflicts of interest. 
} 

\appendixtitles{yes} 
\appendixstart
\appendix
\section[\appendixname~\thesection]{Trapping horizons}
 \label{sec:trapping_horizon}

For a spherically symmetric spacetime, we can generally introduce radial 
null coordinates $\xi_{\pm}$ such that the line element can be written in the 
following double-null form:
\begin{equation}
 ds^{2}=-2 e^{-f(\xi_{+},\xi_{-})}(\xi_{+},\xi_{-})d\xi_{+}d\xi_{-}+R^{2}(\xi_{+},\xi_{-})(d\theta^{2}+\sin^{2}\theta d\phi^{2}).
\end{equation}
We introduce the future-directed radial null vectors $l^{a}_{\pm}\propto (\partial_{\pm })^{a}$ 
such that $g_{ab}l^{a}_{+}l^{b}_{-}=-1$, where $\partial_{\pm}=\partial/\partial\xi_{\pm}$.
Then, we define $\theta_{\pm}$ as 
\begin{equation}
 \theta_{\pm }:=l^{a}_{\pm}\partial_{a}\ln (R^{2}).
\end{equation}

We call a 2-sphere specified with $(\xi_{+},\xi_{-})$ a future (past) trapped sphere if 
$\theta_{+}\theta_{-}>0$ and $\theta_{+}+\theta_{-}< (>)0 $. 
We call a 2-sphere specified with $(\xi_{+},\xi_{-})$ a future (past)
marginal sphere if $\theta_{+}\theta_{-}=0$ and $\theta_{+}+\theta_{-}<(>)0$.
We call a 2-sphere specified with $(\xi_{+},\xi_{-})$ a bifurcating
marginal sphere if $\theta_{+}=\theta_{-}=0$.
We call a 2-sphere specified with $(\xi_{+},\xi_{-})$ an untrapped sphere if 
$\theta_{+}\theta_{-}<0$.
We call a spacetime region a future (past) trapped region, if 
any 2-sphere given by $(\xi_{+},\xi_{-})$ in the region is a future (past) trapped sphere.   
We call a spacetime region an untrapped region, if 
any 2-sphere given by $(\xi_{+},\xi_{-})$ in the region is an untrapped sphere.
We call a hypersurface foliated by future (past) marginal 
spheres a future (past) trapping horizon. 
We call a hypersurface foliated by bifurcating marginal spheres a bifurcating trapping horizon.

See Refs.~\cite{Hayward:1993wb,Hayward:1994bu,Maeda:2009tk} for 
more complete rigorous 
discussions and proofs of the basic properties of trapping horizons.


\bibliography{ref}

\end{document}